\documentclass[aps,prl,superscriptaddress,showpacs,onecolumn,preprint]{revtex4}

\usepackage{graphicx}

\newcommand{\beq}{\begin{equation}}
\newcommand{\eeq}{\end{equation}}
\newcommand{\be}[1]{\begin{equation}\label{#1}}
\newcommand{\ee}{\end{equation}}

\newcommand{\bea}{\begin{eqnarray}}

\newcommand{\eea}{\end{eqnarray}}

\newcommand{\refRef} [1] {Ref.~\cite{#1}}

\newcommand{\refFig} [1] {Fig.~\ref{#1}}

\begin{document}

%\preprint{}

\title{Elasticity of macroscopic isotropic media applied to nano--cantilevers}


\author{N.~S{\o}ndergaard}
\affiliation{Division of Mathematical Physics, LTH, Lunds Universitet, Sweden}
%\email[]{Niels.Sondergaard@matfys.lth.se}

%\thanks{}
\author{S.~Ghatnekar-Nilsson}
\affiliation{Division of Solid State Physics, Lunds Universitet, Sweden}

\author{T.~Guhr}
\affiliation{Division of Mathematical Physics, LTH, Lunds Universitet, Sweden}

\author{L.~Montelius}
\affiliation{Division of Solid State Physics, Lunds Universitet, Sweden}

%\thanks{}


\date{\today}

\begin{abstract}
We analyze recently measured data for the mechanical deflection of
Chromium nano--cantilevers. We show that the deflection curves exhibit
a scaling behavior when applying properly chosen
coordinates. Moreover, the deflection curves are well described by
Euler's theory for elastic, macroscopic rods made of homogenous
material. Apart from the conceptual insight, this also yields a
precise method to deduce the Young's modulus of these
nano--cantilevers. It is found to be considerably smaller than that of
a macroscopic Chromium rod.
\end{abstract}

%\keywords{}

\maketitle

Large research efforts are under way to improve the understanding of
the physics on the nano scale. This is of high interest for basic
research, but also of considerable importance for applications, as the
process continues of making technical devices ever smaller.  When
constructing nano--electro--mechanical systems (NEMS), detailed
knowledge of electronic and mechanical properties is needed.  Most of
the research has been focused on the electronics aspects, while less
is known about the mechanical laws on the nano scale. Nevertheless,
deeper insight into these laws is highly desirable.  An example is
biochemical sensors which are presently being developed. Here, one
aims at accurately measuring very small concentrations of molecules in
gases and liquids by for instance probing the resonance frequency of a
nano--mechanical oscillator as a function of added mass. Thus, precise
knowledge of the elastic properties in those devices is most relevant
for their use as ultra sensitive single molecule detectors.

In this contribution, we address the mechanical deflection of Chromium
nano--cantilevers.  We have three goals: First, we show that the
experimental deflection curves scale when using proper dimensionless
coordinates. Second, we demonstrate that the measured data for the
{\it nano scale} cantilevers can be interpreted using the theory which
Euler developed for {\it macroscopic} rods made of isotropic
material. As we have to deal with relatively large deflections, this
goes considerably beyond a previous analysis valid for small
deflections \cite{sara1,sara2,sara3,lAndl,cleland}. Third, we present
an {\it accurate and quantitative} description of the deflection
curves. This is remarkable, since granularity, cracks or other effects
could modify the mechanical laws on the nano scale. We employ these
insights to deduce most reliable values for the {\it Young's moduli},
i.e.~for the moduli of extension, of these nano--cantilevers.

\begin{figure}[t]                      
\begin{minipage}[l]{0.48\textwidth}
\noindent{\includegraphics[height=4cm]{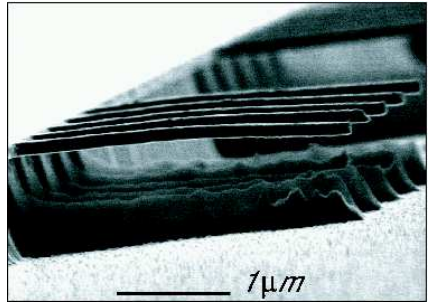}}
\vspace{0.0\textheight} 
\caption{\label{Canti83} Scanning electron microscope picture of the cantilevers of
thickness $t=$ 83~nm.}
\end{minipage}
\end{figure}

The cantilevers were defined by electron beam lithography on a
double--layer resist on a silicon chip \cite{sara1,sara2}. Various
thicknesses of Chromium were evaporated onto the surface. After the
lift--off process, the cantilevers were released from the substrate by
reactive ion etching.  The resulting Chromium cantilevers are $3$
$\mu$m long. The cantilevers are tapered with initial width $200$ nm
and final width $150$ nm at the free end.  Hence, their cross sections
are rectangular having a width of approximately $w=175$~nm and the
thickness $t$ is the same as the evaporated layer measuring 68~nm and
83~nm, respectively. After fabrication, the cantilevers are in a
horizontal position, see \refFig{Canti83} which displays besides the
cantilevers also remnants of the underlying silicon resulting from the
etching step. The latter ridges are disconnected from the cantilever
structure and play no role in the elastic measurements.  
The
cantilevers are mechanically bent by touching them with the tip of an atomic force
microscope (AFM) at 32 points $x$ along the horizontal direction. The
resulting deflections $y$ are measured.  This is done for four
different forces, yielding all together 128 data points for each
cantilever. The actual force $f$ is known up to a constant $f_0$, such
that $f_{\rm meas}=f-f_0$ is the measured force.

Consider a macroscopic rod or beam made of an isotropic material with
Young's modulus $E$. Euler's theory for such a rod comprises certain
approximations to the full Navier equation of elasticity~\cite{lAndl}.
We mention in passing that the theory has also been used in the
statistical physics of rod-like bacteria \cite{ref:bacteria}. We use
coordinates $x$ and $y$ in horizontal and vertical direction,
respectively. The rod is in the $(x,y)$ plane and clamped at one end
at the origin $(0,0)$. Thus, its initial position is uniquely defined
by the angle $\theta_0$ with the vertical. The moment of inertia $I$,
also referred to as bending moment, is measured with respect to the
origin.  A vertical force is applied which bends the rod.  It is
convenient to employ the coordinate $s$ measured along the rod.  The
point where the force acts is $s=l$. Obviously, we have $s=0$ at the
origin.  The bending at a general $s$ is parametrized by the angle
$\theta=\theta(s)$ of the tangent to the rod with the vertical
direction.  We notice $\theta(0)=\theta_0$ and write
$\theta(l)=\theta_l$. By definition the slope of the deflection curve
is given such that $x'(s)=\sin \theta(s)$ and $y'(s)=\cos
\theta(s)$. Thus, \bea x = \int_0^l ds \, \sin \theta(s) \quad {\rm
and} \quad y = \int_0^l ds \, \cos \theta(s) \eea are the coordinates
$(x,y)$ of the point where the force acts on the rod. Hence $y=y(x)$
is the deflection curve. Within his theory, Euler arrives
at~\cite{lAndl} \beq
\label{eq:Pendulum}
\frac{d^2 \theta}{ds^2} = \frac{f}{E I} \, \sin \theta \eeq for the
angle $\theta(s)$. Furthermore, he shows that the torque is given by
$M = E I \theta'(s)$.  Thus, knowledge of bending curves yields
information on $E$. For the present cantilevers the widths change with
$50$ nm over a distance of $3$ $\mu$m. Hence, as a simplifying
assumption we shall assume the width of the cantilevers {\it equal to
their average width}.

The width $w$ and thickness $t$ enter Euler's theory only via the
bending moment given by $I = w t^3/12$ for a rectangular cross
section.  Before solving Eq.~(\ref{eq:Pendulum}), we notice that it
implies a scaling property which suggests the introduction of the {\it
dimensionless} coordinates \beq
\label{scaling}
X = x / \sqrt{EI/f} \quad {\rm and} \quad Y = y / \sqrt{EI/f} \, .
\eeq 
Solutions of Eq.~(\ref{eq:Pendulum}) for constant bending moment
$I$ are given in terms of elliptic integrals as, for example, in the
case of the pendulum equation \cite{goldstein}.  Approximation of the
sine in Eq.~(\ref{eq:Pendulum}) gives a {\it small deflection theory}
of beams. An analysis of cantilevers in the framework of such a theory
has been performed in Refs.~\cite{sara1,sara2,sara3}. Here, however,
we go well beyond this by using the fully fledged theory applying to
the relatively large deflections in the experiment.

An important remark is in order. Large deflections do not necessarily
imply that {\it non--linear} effects have to be taken into account
which would be beyond the regime of {\it Hooke's law}
\cite{antman}. This is no contradiction, as daily experience
illustrates.  A long and thin rod can be bent quite a bit without
losing its elasticity, i.e.~without being permanently deformed. Such a
deformation would be a strong hint for the presence of
non--linearities.  Rather the strains in the rod are sufficiently
small such that linear elasticity still holds. Yet these strains
accumulate to a final large deflection \cite{Love}.

For carbon nanotubes \cite{dresselhaus1} various continuum theories
 have been applied to model the elastic properties. Besides the more
 complicated shell theory  \cite{cntShell} also the Euler theory is
 discussed in \refRef{cntBEAM}.

We now apply Euler's macroscopic theory to our nano--cantilevers.  The
crucial assumption is that the material, i.e.~the Chromium bulk of the
cantilevers, is sufficiently homogenous and isotropic. A priori this
is not obvious, because granularity, cracks or other effects could
destroy the homogeneity and thereby alter the mechanical laws.  The
forces applied by the AFM--tip are much larger than gravity, which we
thus neglect. In particular, we neglect the weight of the part of the
cantilever extending beyond the $x$ position such that the total
length should not matter. At the point of the AFM--tip, the torque is
zero such that $\theta'(l)=0$. With this boundary condition, we
integrate Eq.~(\ref{eq:Pendulum}) once and find \beq
\label{eq:FirstIntegral} \frac{E I}{2 f} \, (\theta'(s))^2 = \cos
\theta_l - \cos \theta(s) \, .  \eeq A second integration yields \beq
\label{eq:XdimL} X = \sqrt{2(\cos \theta_l -\cos \theta_0)} \eeq for
the rescaled position $X$ of the AFM--tip and \bea \label{eq:YdimL} Y
&=& \int_{\theta_l}^{\theta_0} \frac{\cos\theta \,
d\theta}{\sqrt{2(\cos \theta_l -\cos\theta)}} \nonumber \\ &=&
\sqrt{2(\cos \theta_l -\cos \theta_0)} \, \cot \frac{\theta_0 }{2}
\nonumber \\ & & \qquad - 2\,\mathcal{E}(\psi | k) + \mathcal{F}(\psi
|k) \eea for the rescaled deflection $Y$. Here, $\mathcal{F}$ and
$\mathcal{E}$ are the elliptic integrals of the first and second kind,
respectively \cite{abramov}. They depend on an angle defined by $\cos\psi =
\cot(\theta_0/2)\tan(\theta_l/2)$ and on the modulus
$k=\cos(\theta_l/2)$.

To analyze the experimental data, we proceed in two steps. In the
first step, we assume that the cantilevers are in an exact horizontal
position before the force is applied, i.e.~we assume no initial
inclination and set $\theta_0=90^\circ$. The deflection curve depends
thus on two unknown parameters, the elasticity modulus $E$ and the
force offset $f_0$. Fitting the theory to the data yields the values
given in Tab.~\ref{tab:horiz}.
\begin{table}[t] 
\caption{\label{tab:horiz} Elasticity modulus $E$ and force offset $f_0$
for the two cantilevers from the fits assuming no initial inclination.}
\begin{ruledtabular}
\begin{tabular}{l|r|r}  
$t/\mathrm{nm}$ & 68 & 83 \\ \hline 
$E/\mathrm{GPa}$ & 69.5 & 78.0 \\ 
$f_0/ \mathrm{nN}$ & 9.4 & 10.9 \\ 
\end{tabular}
\end{ruledtabular}
\end{table}  
\begin{figure}[t]                      
\begin{minipage}[l]{0.48\textwidth}
\noindent{\includegraphics[height=6cm]{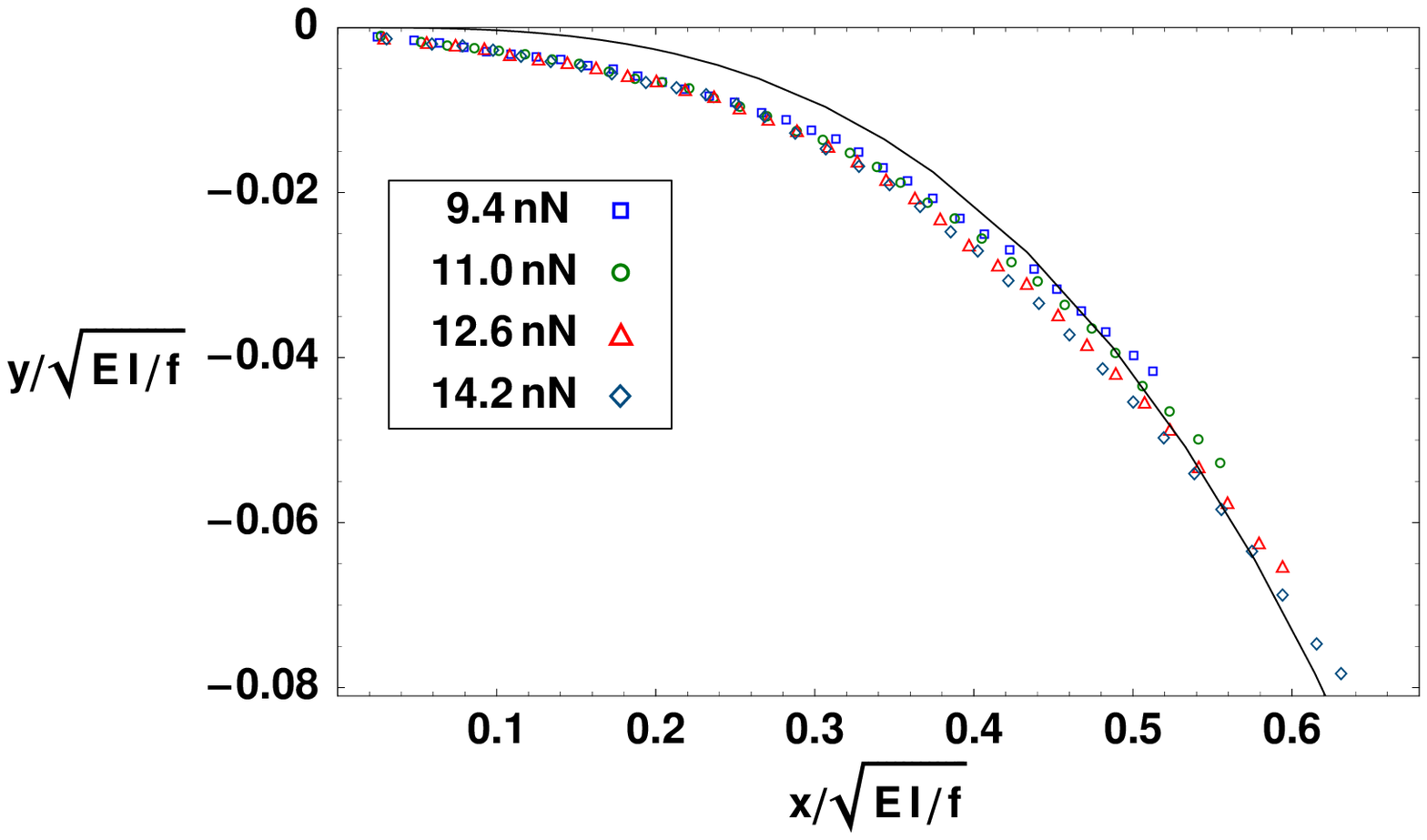}}
\vspace{-0.015\textheight} 
\caption{\label{noSlope2}Fit of Euler's theory (solid line) to the data for the
cantilever with $t=68$~nm assuming no initial inclination. The values
and the symbols for the four different forces are given in the
inlet.}
\end{minipage}
\end{figure} 
\begin{figure}[t]                      
\begin{minipage}[l]{0.48\textwidth}
\noindent{\includegraphics[height=6cm]{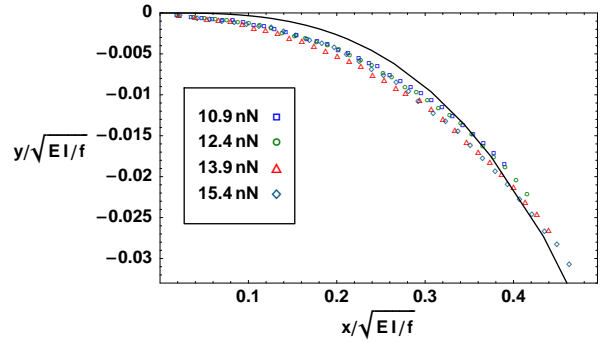}}
\vspace{-0.025\textheight} 
\caption{\label{noSlope3}As Fig.~\ref{noSlope2} for the cantilever with
$t=83$~nm.}
\end{minipage}
\end{figure}
Importantly, the resulting Young's moduli $E$ are about three times
smaller than the macroscopic value of $E_{\rm macro}=248$~GPa. The
order of magnitude is consistent with a previous analysis based on a
small deflection theory \cite{sara1,sara2,sara3}.  The force offsets
$f_0$ depend on the particular AFM but seem consistent with previously
observed offsets in our experimental group.  Our fits are shown in
Figs.~\ref{noSlope2} and~\ref{noSlope3} for the cantilevers with
thickness 68~nm and 83~nm, respectively.   
We notice that the theory gives a qualitative,
but not yet quantitative description of the experiment. Although being
the same function, the theoretical curve in \refFig{noSlope2} and
\refFig{noSlope3} probes different regions in dimensionless
coordinates.  This is mainly due to the difference in thickness
leading to almost twice the bending moment for the thick cantilever.

However, much more important at this point is the observation that the
measured deflections for the four different forces all lie within a
narrow band.  Independently of whether or not the theoretical curve
matches the data, this yields the important insight that the scaling
property~(\ref{scaling}) is realized in the data.
\begin{table}[t] 
\begin{ruledtabular}
\begin{tabular}{l|r|r} 
$t/\mathrm{nm}$ & 68 & 83 \\ \hline 
$E/\mathrm{GPa}$ & 69.5 & 77.6 \\ 
$f_0/ \mathrm{nN}$ & 6.3 & 7.2 \\ 
$\theta_0   \, (\deg)$ & 88.7 & 89.3 \\
$ \chi^2_\nu $  & 1.8 & 1.5\\  
\end{tabular}
\end{ruledtabular}
\caption{ Elasticity modulus $E$, force offset $f_0$,
angle $\theta_0$ and normalized quality $\chi^2_\nu$ for the two
cantilevers from the fits assuming an initial inclination.}
\label{tab:ChiSq1}
\end{table}
\begin{figure}[b]                      
\begin{minipage}[l]{0.48\textwidth}
\noindent{\includegraphics[height=6cm]{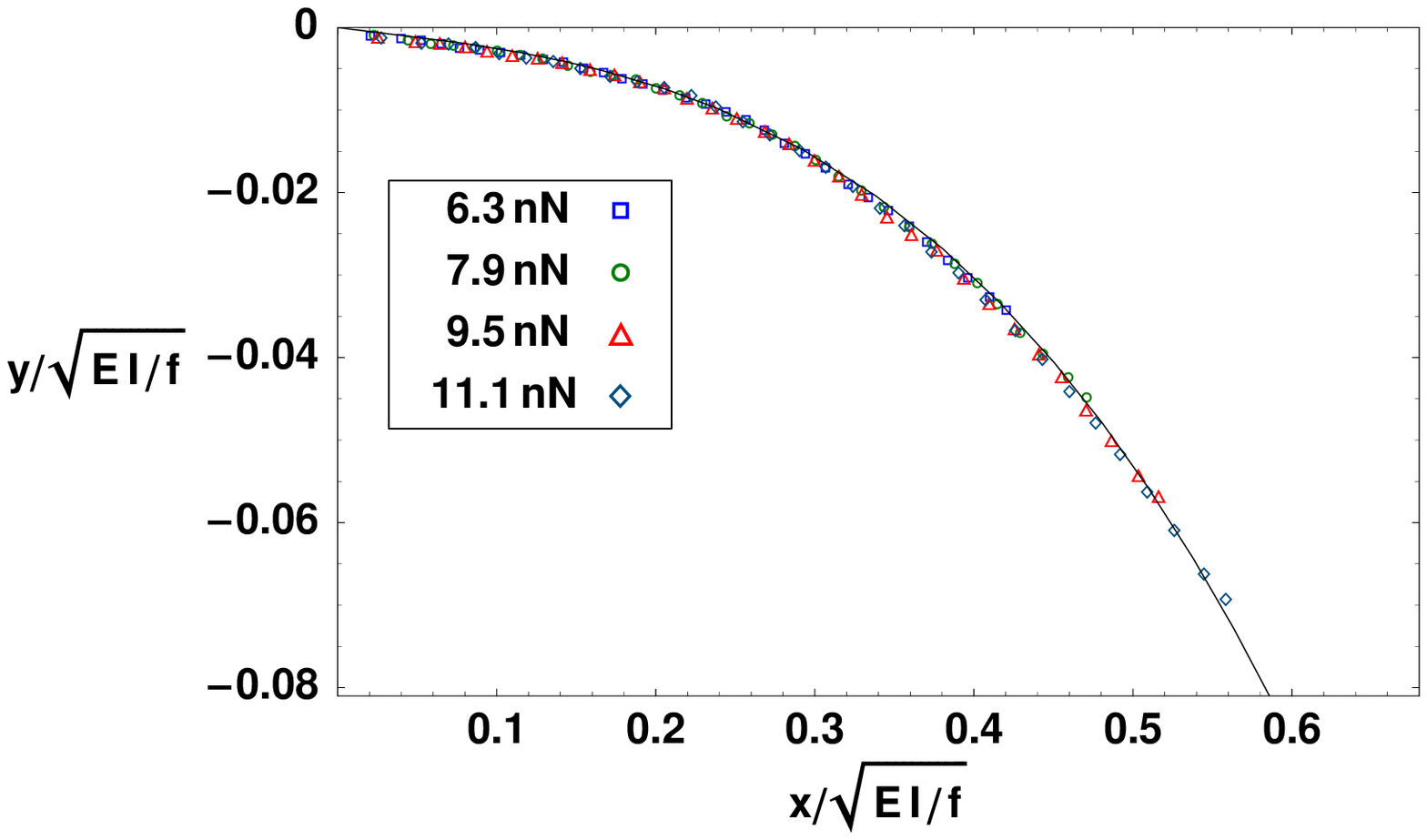}}
\vspace{-0.025\textheight} 
\caption{\label{Dimslope2} Fit of Euler's theory (solid line) to the data for the cantilever with $t=68$~nm assuming an initial inclination. The values
and symbols for the four different forces are given in the
inlet.}
\end{minipage}
\end{figure} 
\begin{figure}[t]                      
\begin{minipage}[l]{0.48\textwidth}
\noindent{\includegraphics[height=6cm]{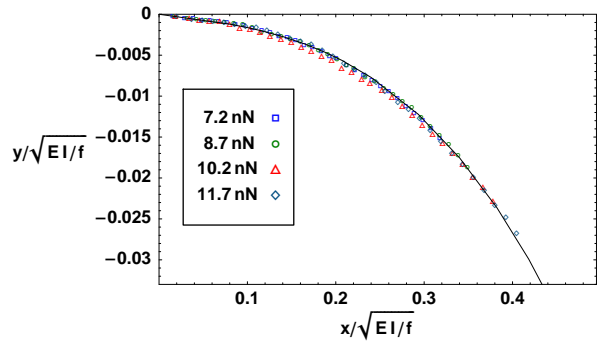}}
\vspace{-0.04\textheight} 
\caption{\label{Dimslope3} As Fig.~\ref{Dimslope2} for the cantilever with
$t=83$~nm.}
\end{minipage}
\end{figure} 
In the second step of our analysis, we try to achieve a quantitative
description of the experiment. A reason for the deviation of the
theoretical curves seen in Figs.~\ref{noSlope2} and~\ref{noSlope3}
could be that length scales other than $\sqrt{EI/f}$ become important,
for example, due to inhomogeneities of the material such as
granularity or small cracks.  However, the presence of the scaling in
the data makes that unlikely. Another possibility would be that
$\sin\theta$ in Eq.~(\ref{eq:Pendulum}) has to be replaced by a more
complicated function due to a general change of the mechanical laws on
the nano scale. We believe, however, that a simpler explanation is
more appropriate. Closer visual inspection of Figs.~\ref{noSlope2}
and~\ref{noSlope3} reveals that the band of the measured points has a
small non--horizontal slope near the origin $(x,y)=(0,0)$. Although
the deflection curve should not be confused with a picture of the bent
cantilever, this indicates an initial, small inclination of the
cantilevers before the force acts.  We can easily take that into
account by employing the angle $\theta_0$ as a fit parameter. Although
we now have to fit three quantities, we will arrive at a consistent
and reliable interpretation. 

The results are given in
Tab.~\ref{tab:ChiSq1}
and the fits are displayed in Figs.~\ref{Dimslope2}
and~\ref{Dimslope3}.  
Excellent agreement between theory and data is
seen. The data points scatter even less than in Figs.~\ref{noSlope2}
and~\ref{noSlope3} due to a slight modification of the Young's moduli
$E$.  It is indeed reassuring that this modification is only
slight. The force offsets $f_0$ change, but within the expected
limits.  It should be noticed that the angles $\theta_0$ are
consistent with our assumption that the initial inclination is only
small.  The main objects of interest, the Young's moduli $E$, are
robust against the addition of the third fit parameter $\theta_0$. We
certainly do not over-fit, as borne out in the values for the
normalized qualities $\chi_\nu^2$ in Tab.~\ref{tab:ChiSq1}.

We turn to an error estimation for our results. In the figures one
sees that the {\it statistical} error is small. Nevertheless, there
are also {\it systematic} errors to be taken into account.  To be
careful, we estimate the error of the fitted Young's moduli $E$ by
Monte Carlo simulations. In these simulations on 1000 synthetic data
sets the uncertainties on thickness, deflection, horizontal position
were set to $2$ nm using normally distributed data. The width,
however, was drawn from a uniform distribution between $150$ nm to
$200$ nm. The uncertainty of the measured force was put to $0.05$ nN
to reflect the uncertainty on the last significant digit.  We arrive
at an error estimate of the order of $9$~GPa for the Young's moduli
$E$.

In conclusion, we have shown that the deflection curves for Chromium
nano--cantilevers scale as predicted by Euler's macroscopic theory and
that the latter yields a quantitative description of the experiment.
Thus, from the data displayed in Figs. 4 and 5, no new scale from
granularity, cracks or other possible material inhomogeneities
appears. We cannot exclude that some effects of microstructure are
present in the Chromium nano--cantilevers that we investigated, but if
so, they do not alter the functional form of the bending law.  We used
this insight to extract Young's moduli $E$ for these nano--cantilevers
from the data. First, the thinner cantilever has a slightly smaller
Young's modulus $E$ than the thicker one. This might indicate a
dependence of the Young's modulus $E$ on the thickness $t$. Such
effects have been seen in the case of crystalline silicon both
experimentally and in molecular dynamics simulations
\cite{experimCanti,atomisticSim}.  At present we cannot confirm a
dependence on thickness for Young's modulus for the two sets of
cantilevers, because the two different Young's moduli $E$ obtained
from the fits are, within the errors, consistent with one single
value.  Nevertheless, we found reliable values which are three times
smaller than the macroscopic value for Chromium. We have no
explanation yet for this drastic reduction. It could be due to
material inhomogeneities \cite{schiotzGrainBulk,arzt1,spaepen1} or
surface to volume effects \cite{dingreville,huangNonlinear} whose
overall influence can be absorbed into the Young's moduli $E$.

However, the situation appears more general as simulations for single
crystal copper nanowires \cite{huangNonlinear} show dependence on
direction and in particular the Young's modulus may increase with
decreasing thickness along certain directions. Still it is plausible,
that some effective theory should be applicable for long and thin
epitaxially grown nanorods \cite{nanowire} and as in the present case
serve to infer the corresponding elasticity modules from
experiments. That theory may have to include anisotropy but will still
be in the spirit of Euler's theory.

\begin{acknowledgments}
We thank C.~Ellegaard and M.~Oxborrow for fruitful discussions.  We
acknowledge financial support from the Swedish Research Council, the Swedish Foundation for Strategic Research and the EU-IST-2001-33068 project Nanomass II.
\end{acknowledgments}

\end{document}